\documentstyle[11pt,fleqn]{article}
\newcommand{\R}{\mbox{$I\!\!R$}}             %%% real numbers
\newcommand{\C}{\mbox{$I\!\!\!\!C$}}         %%% complex numbers

\begin{document}

%%%%%%%%%%%%%%  Preprint number %%%%%%%%%%%%%%%%%%%%%%%%%%

\hfill{\sl preprint -  }
\par
\bigskip
\par
\rm

%%%%%%%%%%%%%   Title %%%%%%%%%%%%%%%%%%%%%%%%%%

\par
\bigskip
\LARGE
\noindent
{\bf A review on recent results of the $\zeta$-function
regularization procedure in curved spacetime}
\par
\bigskip
\par
\rm
\normalsize

%%%%%%%%%%%%%%%%%%%%%%%%%%%%%%%%%%%%%%%%%%%%%

%%%%%%%%%%%% Author %%%%%%%%%%%%%%%%%%%%%%%%%%%

\large
\noindent {\bf Valter Moretti}

\large
\smallskip

\noindent
Department of Mathematics, Trento University and
Istituto Nazionale di Fisica Nucleare,\\
Gruppo Collegato di Trento,
I-38050 Povo (TN), Italy.\\
E-mail: moretti@science.unitn.it

\large
\smallskip

\rm\normalsize

%%%%%%%%%%%%%%%%%%%%%%%%%%%%%%%%%%%%%%%%%%%

%%%%%%%%%%%% Date %%%%%%%%%%%%%%%%%%%%%%%%%%

\par
\bigskip
\par
\hfill{\sl February 1999}
\par
\medskip
\par\rm

%%%%%%%%%%%%%%%%%%%%%%%%%%%%%%%%%%%%%%%%%%%

%%%%%%%%%%%% Abstract %%%%%%%%%%%%%%%%%%%%%%%%

\noindent
{\bf Abstract:}
Some recent (1997-1998) theoretical results concerning the 
$\zeta$-function regularization procedure used to renormalize, at one-loop, 
the effective Lagrangian, the field fluctuations and the stress-tensor and 
some applications are reviewed.
\par

 \rm
%%%%%%%%%%%%%%%%%%%%%%%%%%%%%%%%%%%%%%%%%%%%%%%%

%%%%%%%%%%%%% PACS numbers %%%%%%%%%%%%%%%%%%%%%%%%%
%\noindent{\sl PACS number(s):\hspace{0.3cm}
%04.62.+v, 04.70.Dy}
%\par
%\bigskip
%\rm

%%%%%%%%%%%%%%%%%%%%%%%%%%%%%%%%%%%%%%%%%%%%%%%%

%%%%%%%%%%%%%%%%  Article text %%%%%%%%%%%%%%%%%%%%%%%

\section{Quasifree QFT in curved static manifolds, Euclidean approach
$\zeta$-function technique.}

\noindent {\bf 1.1} {\em Preliminaries.} Let $(L,g^L_{\mu\nu})$ be 
a Lorentzian manifold which is supposed to be {\em globally hyperbolic}.
This manifold is said (locally) {\em static} whenever it admits a (local)
 time-like Killing vector $\partial_t$  normal to a Cauchy surface 
$\Sigma$. In other words, the manifold admits a (local) coordinate frame
$(x^0,x^1,x^2,x^3)\equiv (t,\vec{x})$,
where $g_{0i} = 0$ ($i=1,2,3$) and $\partial_t g_{\mu\nu}=0$. 
Let us consider a real scalar field $\phi$ propagating in $L$, its evolution 
equation can be written down
\begin{eqnarray}
A'_L \phi = 0 \label{evolution}\:,
\end{eqnarray}
where $A'_L := -\nabla_\mu\nabla^\mu + V$, $V$ being a smooth scalar 
field of the form
\begin{eqnarray}
V(x) := \xi R+m^2 +V'(x) \label{V}\:.
\end{eqnarray}
Above, $V'$ is another smooth scalar field satisfying $\partial_t V=0$,
 moreover $\xi$ is a constant
$\xi \in \R$, $R$ is the scalar 
curvature and $m^2$ the squared mass of the particles associated to the field.
The operator $A'_L$ works on a space of real-valued $C^\infty$ functions.
A straightforward way to build up the QFT is the definition of opportune
Green functions of the operator $A'_L$ \cite{fr}, 
in particular the Feynman propagator
$G_F(x,x')$ or, equivalently, the Wightman functions $W_{\pm}(x,x')$. Then, 
generalizations of GNS theorems \cite{kw} in curved spacetime 
allow one to build up a corresponding Fock space
and a quasifree QFT, $G_F$ corresponding to a (not necessarily pure) 
vacuum state. 
In a globally hyperbolic region where the static 
coordinates above are defined, it must be possible a  
choice of Green functions which are  invariant under translations of
 the local static Killing time $x^0=t$. These Green 
functions should determine
a static vacuum which may represent a thermal state. Moreover
these coordinates  allow
one to perform the Wick rotation to get the Euclidean formulation of the QFT.
This means that (locally) one can pass from the Lorentzian manifold
$(L,g^L_{\mu\nu})$ to a Riemannian manifold $(M, g_{ab})$ by the analytic
continuation $t\to i\tau$ where $t,\tau \in \R$. This defines a (local) Killing
vector $\partial_\tau$ in the Riemannian manifold and a corresponding
(local) ``static'' coordinate frame $(\tau,\vec{x})$ therein. As is well-known 
\cite{fr},
in the case the Riemannian manifold has been made compact along 
the Euclidean
time $\tau$ with a period $\beta$, $T=1/\beta$ has to be interpreted as 
the temperature of the quantum state. The corresponding Feynman propagator,
 whenever 
continued in the Riemannian manifold, admits $\beta$ as Euclidean temporal 
period.
In this approach, the Feynman propagator $G_F(t-t',\vec{x},\vec{x}')$
 determines and, generally speaking \cite{fr}, itself 
is completely determined by,
 a proper Green function (in the spectral theory
sense) $S_\beta(\tau-\tau',\vec{x},\vec{x}')$
of a self-adjoint extension $A$ of the operator
\begin{eqnarray}
A' :=  -\nabla_a\nabla^a + V(\vec{x}) \::\:  C_0^\infty(M) \to L^2(M,d\mu_g)\:.
\end{eqnarray}
(Above, $M$ can be restricted to an opportune  region where the metric
is static.) 
$S_\beta(\tau-\tau',\vec{x},\vec{x}')$,
 said the Schwinger function, is the integral kernel of
$A^{-1}$, supposing $A>0$.

The {\em partition function}
 of the quantum state can be computed as the functional
integral evaluated over the field configurations periodic with period $\beta$
in the Euclidean time
\begin{eqnarray}
Z_\beta = \int {\cal D}\phi \:e^{-S_E[\phi]}\:,
\end{eqnarray}
the Euclidean action $S_E$ being ($d\mu_g := \sqrt{g} d^4x$)
\begin{eqnarray}
S_E[\phi] = \frac{1}{2}\int_M d\mu_g(x) \:\phi(x) A_x \phi(x)\:.
\end{eqnarray}
Thus, 
formally speaking, one has 
\begin{eqnarray}
Z_\beta = \left\{ 
\det \left(\frac{A}{\mu^2}\right)\right\}^{-1/2}\:,\label{det}
\end{eqnarray}
where $\mu$ is a mass scale which is necessary for dimensional reasons.\\

\noindent {\bf 1.2} {\em The $\zeta$-function technique.} The most intriguing
problem related to the applications of  (\ref{det})
concerns the interpretation of the determinant of an operator.  
A interesting suggestion is given by the $\zeta$-function procedure.
Suppose $A$ is a $n\times n$ positive-definite Hermitian matrix with 
eigenvalues $0<\lambda_1\leq \lambda_2\leq \cdots \leq \lambda_n$.
Then one can define the complex-valued function
\begin{eqnarray}
\zeta(s|A) = \sum_{j=1}^n \lambda_j^{-s}\:,
\end{eqnarray}
where $s\in \C$. (Notice that $\lambda_j^{-s}$ is well-defined since 
$\lambda_j>0$.)
It is now a trivial task to prove that 
\begin{eqnarray}
\det A = e^{-\frac{d\zeta(s|A)}{ds}|_{s=0}}\:. \label{det'}
\end{eqnarray}
In the case $M$ is a $D$-dimensional Riemannian compact manifold 
and $A'$ is bounded below by some constant 
$b\geq 0$, this procedure can be generalized to operators.
In this case, $A'$ admits the Friedrichs self-adjoint extension $A$
which is also bounded below by the same bound of $A'$, moreover the spectrum
of $A$ is discrete and each eigenspace has a finite dimension.
Then one can consider the series with $s\in \C$ (the prime on the 
sum means that any possible null eigenvalues is omitted)
\begin{eqnarray}
\zeta(s|A/\mu^2) := {\sum_j}' \left( \frac{\lambda_j}{\mu^2}\right)^{-s}\:.
\label{sum}
\end{eqnarray}
As is well-known, the series above converges provided  $Re$ $s >D/2$, where
$D$ is the dimension of $M$. Moreover, it is possible to continue the
 right-hand side above into a meromorphic function of $s$ which is regular
at $s=0$ (see \cite{m1} for a short review of the principal properties 
of the convergence of the series above and for the corresponding 
bibliography). 
Following (\ref{det}) and (\ref{det'}), the idea \cite{haw,e-z} is to define
\begin{eqnarray}
Z_\beta := e^{\frac{1}{2}\frac{d\:}{ds}|_{s=0}  \zeta(s|A/\mu^2)}\:, 
\label{zeta}
\end{eqnarray}
where  the function $\zeta$ on the right-hand side is the analytic continuation
of that defined in (\ref{sum}). 
It is possible the define the $\zeta$ function in terms of the heat kernel
of the operator $A$, $K(t,x,y|A)$. This is the smooth integral kernel of
the (Hilbert-Schmidt, trace-class) operators
$e^{-tA}$, $t>0$. One has, for $Re$ $s>D/2$,
\begin{eqnarray}
\zeta(s|A/\mu^2) = \int_M d\mu_g(x) \int_0^{+\infty} 
dt\:\frac{\mu^{2s}t^{s-1}}{\Gamma(s)} \left[ K(t,x,x|A) -P_0(x,x|A) \right]
\:, \label{*}  
\end{eqnarray}
$P(x,y|A)$ is the integral kernel of the projector on the null-eigenvalues
eigenspace of $A$.\\
If the manifold $M$ is not compact, $A$
has also a  continuous-spectrum part, however, 
it is still possible to generalize the 
definitions and the results above considering opportune integrals on the 
spectrum of the operator $A$ provided $A$ is strictly positive
(e.g, see \cite{wald79}).\\
Another very useful tool is the {\em local} $\zeta$ function \cite{wald79}, 
which can be 
defined in two different but equivalent ways:
\begin{eqnarray}
\zeta(s,x|A/\mu^2) =  \int_0^{+\infty} 
dt\:\frac{\mu^{2s}t^{s-1}}{\Gamma(s)} \left[ K(t,x,x|A) -P_0(x,x|A) \right]
\:, \label{**}  
\end{eqnarray}
and, $\phi_j$ being the smooth eigenvector of the eigenvalue $\lambda_j$,
\begin{eqnarray}
\zeta(s,x|A/\mu^2) = {\sum_j}' \left(\frac{\lambda_j}{\mu^2}\right)^{-s}
\phi_j(x)\phi^*_j(x)\:. \label{***}
\end{eqnarray}
Both the integral and the series converges for $Re$ $s>D/2$. The local
zeta function  enjoys the same analyticity properties of the integrated
$\zeta$ function (\cite{wald79,m1}). For future convenience it is also 
useful to define, in the sense of the analytic continuation, 
\begin{eqnarray}
\zeta(s,x,y|A/\mu^2) =  \int_0^{+\infty} 
dt\:\frac{\mu^{2s}t^{s-1}}{\Gamma(s)} \left[ K(t,x,y|A) -P_0(x,y|A) \right]
\: \label{**'}  
\end{eqnarray}
(see \cite{m1,m2} for the properties of this off-diagonal $\zeta$-function).
In the 
framework of the $\zeta$-function regularization framework, 
the {\em effective Lagrangian} is defined as
\begin{eqnarray}
{\cal L}(x|A)_{\mu^2}
:= \frac{1}{2}\frac{d\:}{ds}|_{s=0}  \zeta(s,x|A/\mu^2)\:, 
\label{L}
\end{eqnarray}
and thus, in a thermal theory,
 $Z_\beta = e^{-S_\beta}$ where $S_\beta = \int d\mu_g 
{\cal L}_{\beta\mu^2}$.
A first recent 
result which, in compact manifolds at least, generalizes to any 
dimension an earlier results by Wald \cite{wald79}, has
been obtained in \cite{m1}. This result proves how the effective Lagrangian 
can be obtained by a {\em point-splitting procedure} (see below);
for $D$ even it reads
\begin{eqnarray}
{\cal L}(y|A)_{\mu^2} &=& \lim_{x\to y}
\left\{ - \int_0^{+\infty}\frac{dt}{2t}K(t,x,y|A) - 
\frac{a_{D/2}(x,y)}{2(4\pi)^{D/2}}\ln\frac{\mu^2\sigma(x,y)}{2} \right.
\nonumber\\
&+& \left.  \sum_{j=0}^{D/2-1} (\frac{D}{2}-j-1)! \frac{a_j(x,y|A)}
{2(4\pi)^{D/2}} \left(\frac{2}{\sigma(x,y)} \right)^{D/2-j}\right\}
- 2\gamma \frac{a_{D/2}(y,y)}{2(4\pi)^{D/2}}\:,
\end{eqnarray}
and for $D$ odd (notice that $\mu$ disappears from the final result)
\begin{eqnarray}
{\cal L}(y|A)_{\mu^2} &=& \lim_{x\to y}
\left\{ - \int_0^{+\infty}\frac{dt}{2t}K(t,x,y|A) - 
\sqrt{\frac{2}{\sigma(x,y)}}\frac{a_{(D-1)/2}(x,y)}{2(4\pi)^{D/2}} \right.
\nonumber\\
&+& \left.  \sum_{j=0}^{(D-3)/2} \frac{(D-2j-2)!!}{2^{(D+1)/2-j}} 
\frac{a_j(x,y|A)}
{2(4\pi)^{D/2}} \left(\frac{2}{\sigma(x,y)} \right)^{D/2-j}\right\}\:.
\end{eqnarray}
Above, $\sigma(x,y)$ is one half the square of the geodesical distance of
$x$ from $y$ and the coefficients $a_j$ are the well-known off-diagonal
coefficients of the small-$t$ expansion of the heat-kernel (see \cite{m1}
for a short review on the properties of these coefficients).\\

\section{Improvements of the local $\zeta$-function technique.}

\noindent {\bf 2.1} {\em Generalizations of the local $\zeta$ function 
technique.} Besides the effective Lagrangian and the effective action,
further important one-loop quantities are the {\em (quantum) field fluctuation}
and the {\em averaged (quantum) stress tensor}. These quantities are given, in 
terms of the Euclidean path integral, by
\begin{eqnarray}
<\phi^2(x)> &=& \frac{\delta}{\delta J(x)}|_{J\equiv 0}
\ln \int {\cal D} \phi\: e^{-S_E + \int d\mu_g \phi^2 J} \label{phi}\:,\\
<T_{ab}(x)> &=& \frac{2}{\sqrt{g(x)}}\frac{\delta}{\delta g^{ab}(x)}
\ln \int {\cal D} \phi\: e^{-S_E[g]} \:. \label{tab}
\end{eqnarray}
A very popular method to compute the quantities above which in the practice
diverge, is the so-called {\em point-splitting procedure} \cite{bd,fu,waldlibro},
anyhow, it is possible to generalize the $\zeta$-function method in order to
build up opportune $\zeta$ functions which regularize  the 
quantities above directly, 
similarly to the procedure for the effective Lagrangian
\cite{m0,im,m1,m2}.  Let us consider the stress tensor, 
the way to get a direct $\zeta$-function regularization procedure is based on
the following chain of formal identities \cite{m0}
\begin{eqnarray}
& &{\sqrt{g(x)}} <T_{ab}(x)> \mbox{``}=\mbox{''} 
2\frac{\delta}{\delta g^{ab}(x)}
\ln Z_\beta \mbox{``}=\mbox{''}  \frac{\delta}{\delta g^{ab}(x)}
\frac{d}{ds}|_{s=0} 
\zeta(s|A/\mu^2) \nonumber\\ 
& & \mbox{``}=\mbox{''} \frac{\delta}{\delta g^{ab}(x)}\frac{d}{ds}|_{s=0} 
{\sum_j}'\left(\frac{\lambda_j}{\mu^2} \right)^{-s} \mbox{``}=\mbox{''} 
\frac{d}{ds}|_{s=0} \mu^{-2s}
{\sum_j}' \frac{\delta \lambda_j^{-s}}{\delta g^{ab}(x)}\:. 
\end{eqnarray}
Thus, one {\em define} the $\zeta$-regularized (or renormalized) stress tensor
as 
\begin{eqnarray}
<T_{ab}(x|A)>_{\mu^2}:=
\frac{1}{2} \frac{d}{ds}|_{s=0} Z_{ab}(s,x|A/\mu^2) \label{zetatab}\:,
\end{eqnarray}
where, {\em in the sense of the analytic continuation of the left-hand side}
\begin{eqnarray}
Z_{ab}(s,x|A/\mu^2) := 2{\sum_j}'\mu^{-2s}
 \frac{\delta \lambda_j^{-s}}{\delta g^{ab}(x)}\:. 
\label{zetab}
\end{eqnarray}
The mathematical problem is whether the right-hand side above can be  computed
in the practice and whether it defines
 an analytic function of $s$ in a neighborhood of $s=0$.  
We have the result \cite{m0,m2}\\

\noindent{\bf T1.} 
{\em If $M$ is compact, $A\geq 0$ and $\mu^2>0$, then $Z_{ab}(s,x|A/\mu^2)$ is 
well-defined and is a $C^\infty$ function of $x$ which is also meromorphic 
in $s\in \C$. In particular, it is analytic in a neighborhood of $s=0$.}\\

\noindent In the practice, the result above has been checked also in several 
noncompact manifolds also containing singularities \cite{m0}. In these cases
the summation on the right hand side of (\ref{zetab}) has to be changed
into a spectral integration. The form of the series on the right-hand side 
of (\ref{zetab}) is \cite{m0,m2},
\begin{eqnarray}
s{\sum_j}' \left\{ \frac{2}{\mu^2}\left(\frac{\lambda_j}{\mu^2}
\right)^{-s-1} 
T_{ab}[\phi_j,\phi^*_j](x)
+ g_{ab}(x) \left(\frac{\lambda_j}{\mu^2}\right)^{-s}\right\} \:.\nonumber
\end{eqnarray}
Above $T_{ab}[\phi_j,\phi^*_j](x)$ is the classical stress tensor
evaluated on the modes of the Euclidean-motion operator
$A = -\Delta +\xi R + m^2 +V(x)'$ (see \cite{m0,m2} for details). The series
converges for $Re$ $s>3D/2 +2$. 

Similarly, it is possible to built up a $\zeta$ function for the field 
fluctuation \cite{im,m1}. One has
\begin{eqnarray}
<\phi^2(x|A)>_{\mu^2} := \frac{d}{ds}|_{s=0} \Phi(s,x|A/\mu^2)\:, \nonumber 
\end{eqnarray}
where 
\begin{eqnarray}
\Phi(s,x|A/\mu^2)    := \frac{s}{\mu^2} \zeta(s+1,x|A/\mu^2)
\:.
\end{eqnarray}
The properties of these functions have been studied in \cite{im,m1} and
several applications on concrete cases are considered ({\em e.g.}
cosmic-string spacetime and homogeneous spacetimes). In particular, in 
\cite{m1}, the problem of the change of the parameter $m^2$ in the field
fluctuations has been studied.\\

\noindent {\bf 2.2} {\em Physically correctness of the given regularization
procedures.} We are now concerned with the physical interest of the found 
regularization techniques. To this end, the following quite general 
results are relevant
and prove that the proposed technique are physically good candidates
\cite{m0,m2}\\

\noindent {\bf T2}
{\em If $M$ is compact, $A\geq 0$ and $\mu^2>0$, and the averaged 
quantities above are those defined above in terms of local $\zeta$-function
regularization, then}

(a) {\em $<T_{ab}(x|A)>_{\mu^2}$ 
defines a $C^\infty$ symmetric tensorial field.} 

(b) {\em Similarly to the classical result,}
\begin{eqnarray} 
\nabla^b <T_{bc}(x|A)>_{\mu^2} = -\frac{1}{2} <\phi^2(x|A)>_{\mu^2}
 \nabla_c V'(x)\:.
\end{eqnarray}

(c)  {\em Concerning the trace of the stress tensor, it is naturally decomposed
in the classical and the (correct) anomalous part}
\begin{eqnarray}
g^{ab}(x)<T_{ab}(x|A) >_{\mu^2} &=& \left(\frac{\xi_D-\xi}{4\xi_D-1}\Delta 
-m^2 -V'(x) \right)<\phi^2(x|A)>_{\mu^2}\nonumber\\
&+& \delta_D \frac{a_{D/2}(x,x|A)}{(4\pi)^{D/2}}  - P_0(x,x|A)\:,
\end{eqnarray}
 {\em where $\delta_D =0$ if $D$ is 
odd and $\delta_D=1$ if $D$ is even, $\xi_D = (D-2)/[4(D-1)]$.} 

(d) {\em for any $\alpha>0 $}
\begin{eqnarray}
<T_{ab}(x|A)>_{\alpha\mu^2} = <T_{ab}(x|A)>_{\mu^2} + t_{ab}(x)\ln \alpha\:,
\end{eqnarray}
{\em where, the form of $t_{ab}(x)$ which depends on the geometry only and 
 is in agreement with Wald's axioms} \cite{waldlibro}, {\em has been given
in} \cite{m2}.

(e) {\em In the case $\partial_0 = \partial_\tau$ is a global Killing vector,  
the manifold admits periodicity $\beta$ along the lines tangent to 
$\partial_0$ and $M$ remains smooth (near any fixed points of the
 Killing orbits) 
fixing arbitrarily $\beta$ in a 
neighborhood and, finally, $\Sigma$ is a global surface everywhere 
normal to $\partial_0$, then}
\begin{eqnarray}
\frac{\partial\:\:}{\partial \beta} \ln Z(\beta)_{\mu^2} =  
\int_\Sigma d\vec{x} \sqrt{g(\vec{x})}
 <T_0^0(x,\beta|A)>_{\mu^2}\:. 
\end{eqnarray}\\

\noindent Another general result, concerning the possibility to get 
a Lorentzian theory from an Euclidean one, is the following one \cite{m2}.\\

\noindent{\bf T3.}
{\em Let  $M$ be compact, $A\geq 0$, $\mu^2>0$, let $M$ be  also globally 
static with global Killing time $\partial_\tau$ and (orthogonal) global spatial
section $\Sigma$ and finally, $\partial_\tau V'\equiv 0$. Then} 

(a) $\partial_\tau <\phi^2(x|A)>_{\mu^2} \equiv 0$;

(b) $\partial_\tau {\cal L}(x|A)_{\mu^2} \equiv 0$;

(c) $\partial_\tau <T_{ab}(x|A)>_{\mu^2}   \equiv 0$;

(d) $ <T_{0i}(x|A)>_{\mu^2}   \equiv 0$ {\em for} $i=1,2,3,...,D-1$

{\em where
the averaged  quantities above are those defined above in terms of local 
$\zeta$-function regularization and coordinates $\tau \equiv x^0,
\vec{x}\in \Sigma$  are employed.}\\

\noindent These properties allow one to continue 
the Euclidean considered quantities  into imaginary values of 
the coordinate $\tau \mapsto it$ obtaining {\em real } functions of the
Lorentzian time $t$.

Some of the properties above (concerning {\bf T1},{\bf T2}, {\bf T3}) 
have been checked also in noncompact 
and symmetric manifolds (Rindler spacetime, cosmic string spacetime, 
Einstein's open spacetime, $H^N$ spaces, G\"{o}del spacetime, BTZ spacetime)
\cite{m0,im,caldarelli,radu,bmvz}. In particular,
 the theory of the regularization of the
stress tensor and field fluctuations  via local $\zeta$ function has been 
successfully employed to compute the back reaction on the three-dimensional
 BTZ metric  \cite{bmvz} in the case of the singular ground state containing
 a naked singularity and a semiclassical implementation of the cosmic
 censorship conjecture has been found.\\

\noindent {\bf 2.3.} {\em The relation with the point-splitting technique.}
The procedure of the point-splitting to renormalize the field 
fluctuation as well as the  stress tensor we are concerned with
\cite{bd,waldlibro}
 can be summarized as
\begin{eqnarray}
<\phi^2(y)>_{\scriptsize \mbox{ps}} 
&=& \lim_{x\to y} \left\{ G(x,y) - H(x,y) \right\} 
+ g_{ab}(y) Q(y)
\label{ps1}\:,\\
<T_{ab}(y)>_{\scriptsize \mbox{ps}} &=& \lim_{x\to y} {\cal D}_{ab}(x,y)
\left\{ G(x,y) - H(x,y) \right\} 
+ g_{ab}(y) Q(y)\:,\label{ps2}
\end{eqnarray}
where, $G(x,y)$ is one half
 the Hadamard function ({\em i.e.},
one half the sum of the two Wightman function) of the considered quantum state
or, in Euclidean approach, the corresponding Green (Schwinger) function.
$H(x,y)$ is the {\em Hadamard local fundamental solution},
 a {\em parametrix} for
the Green function which depends on the {\em local} 
geometry only and takes the short-distance singularity  into account. 
$H(x,y)$ is represented in terms of a truncated series of functions of 
$\sigma(x,y)$.
The operator ${\cal D}_{ab}(x,y)$ is a bi-tensorial operator obtained by
``splitting'' the argument of the classical expression of the stress tensor
(see \cite{m2} for a quite general expression of this operator). Finally
$Q(y)$ is a scalar obtained by imposing several physical 
conditions (essentially, the appearance of the conformal anomaly,
the conservation of the stress tensor and the
 triviality of the Minkowskian limit)
\cite{waldlibro} on the
 left-hand side of (\ref{ps2}) (see 
\cite{bd,fu,waldlibro,m2} for details). 
The expression of $H(x,y)$ is the following one,
in a geodesically convex neighborhood containing both $x$ and $y$,
\begin{eqnarray}
H(x,y) &=& \frac{\sum_{j=0}^L u_j(x,y) \sigma(x,y)^j}{(4\pi)^{D/2}
(\sigma(x,y)/2)^{D/2-1}} + \delta_D \left[\sum_{j=0}^{M} v_j(x,y) 
\sigma(x,y)^j \ln \left(\frac{\sigma(x,y)}{2}\right)\right] \nonumber\\
&+& \delta_D \sum_{j=0}^{N} w_j(x,y) \sigma(x,y)^j \label{hadamard}\:.
\end{eqnarray}
Above, $L,M,N$ are fixed integers (see \cite{m2} for details), 
$\delta_D=0$ if $D$ is odd and $\delta_D=1$ otherwise. The coefficients
$u_j$ and $v_j$ are smooth functions of $(x,y)$ which are completely 
determined by the local 
geometry. Conversely, the coefficients $w_j$ are determined
once one has fixed $w_0$. Dealing with Euclidean approaches, it is possible 
to explicit $u_j$ and $v_j$ in terms of heat-kernel coefficients \cite{m1,m2}.
The problem is the determination  of the coefficient $w_0$. This coefficient
should determine the form of the term $Q$ above. However, there is no guarantee
that {\em any}
 choice of $w_0$ determine such a  $Q$ that the obtained stress tensor 
fulfills the requested physical conditions. (Anyhow, as well-know
 \cite{waldlibro} a possible  choice is $w_0\equiv 0$).\\

\noindent One has the following result \cite{m1,m2}.

\noindent {\bf T4.}
{\em If $M$ is compact, $A\geq 0$ and $\mu^2>0$, and the averaged 
quantities above are those defined above in terms of local $\zeta$-function
regularization, then}
\begin{eqnarray}
<\phi^2(y|A)>_{\mu^2} &=& \lim_{x\to y} \left\{ G(x,y) - H(x,y) \right\} 
\label{ps1*}\:,\\
<T_{ab}(y|A)>_{\mu^2} &=& \lim_{x\to y} {\cal D}_{ab}(x,y)
\left\{ G(x,y) - H_{\mu^2}(x,y) \right\} + g_{ab}(y) Q(y)
\label{ps2*}\:,
\end{eqnarray}
{\em where $G(x,y) = \zeta(1,x,y|A/\mu^2)$ given in} (\ref{**'}),
{\em $H_{\mu^2}$ is completely determined by posing}
\begin{eqnarray}
w_0(x,y) := -\frac{a_{D/2-1}(x,y|A)}{(4\pi)^{D/2}}[2\gamma + \ln \mu^2]\:,
\end{eqnarray}
{\em and the term $Q$ is found to be}
\begin{eqnarray}
D Q(y) = -P_0(y,y|A) +\delta_D \frac{a_{D/2}(y,y|A)}{(4\pi)^{D/2}} \:.
\end{eqnarray}\\

\noindent All the requests on the stress tensor necessary to determine
$Q$ in the point-splitting  approach are now fulfilled by the left-hand
side of (\ref{ps2*}). Moreover, not depending on the $\zeta$-function
approach, the choice above for $w_0$ and $Q$ should work also in Lorentzian
and noncompact manifolds as pointed-out in \cite{m2}, where this conjecture
has been checked for Minkowski spacetime.\\

It is worthwhile stressing that, in 1997-1998, several results have been 
obtained concerning the so-called {\em multiplicative anomaly} via 
$\zeta$-function techniques \cite{mult}. Anyhow, we shall not review 
these results here because these are quite far from the main arguments
of this talk.

\end{document}